\documentclass[conference]{IEEEtran}
\IEEEoverridecommandlockouts
\usepackage{cite}
\usepackage{amsmath,amssymb,amsfonts}
\usepackage{algorithmic}
\usepackage{graphicx}
\usepackage{textcomp}
\usepackage{xcolor}
\usepackage{subfig}
\usepackage[inline]{enumitem}
\usepackage{relsize}
\graphicspath{ {./images/} }

\def\BibTeX{{\rm B\kern-.05em{\sc i\kern-.025em b}\kern-.08em
    T\kern-.1667em\lower.7ex\hbox{E}\kern-.125emX}}
\begin{document}

\title{A Machine Learning Pipeline to Examine Political Bias with Congressional Speeches
}



\author{\IEEEauthorblockN{Prasad Hajare\IEEEauthorrefmark{1},
Sadia Kamal\IEEEauthorrefmark{2}, Siddharth Krishnan\IEEEauthorrefmark{3} and
Arunkumar Bagavathi\IEEEauthorrefmark{4}}
\IEEEauthorblockA{Department of Computer Science\\
Oklahoma State University, \IEEEauthorrefmark{3}UNC Charlotte\\
Email: \IEEEauthorrefmark{1}phajare@okstate.edu,
\IEEEauthorrefmark{2}sadia.kamal@okstate.edu,
\IEEEauthorrefmark{3}skrishnan@uncc.edu,
\IEEEauthorrefmark{4}abagava@okstate.edu}}

\maketitle

\begin{abstract}
Computational methods to model political bias in social media involve several challenges due to heterogeneity, high-dimensionality, multiple modalities, and the scale of the data. Political bias in social media has been studied in multiple viewpoints like media bias, political ideology, echo chambers, and controversies using machine learning pipelines. Most of the current methods rely heavily on the manually-labeled ground-truth data for the underlying political bias prediction tasks. Limitations of such methods include human-intensive labeling, labels related to only a specific problem, and the inability to determine the near future bias state of a social media conversation. In this work, we address such problems and give machine learning approaches to study political bias in two ideologically diverse social media forums: \emph{Gab} and \emph{Twitter} without the availability of human-annotated data. Our proposed methods exploit the use of transcripts collected from political speeches in US congress to label the data and achieve the highest accuracy of $70.5\%$ and $65.1\%$ in Twitter and Gab data respectively to predict political bias. We also present a machine learning approach that combines features from cascades and text to forecast cascade's political bias with an accuracy of about $85\%$. 
\end{abstract}

\begin{IEEEkeywords}
Political bias, predictive analytics, bias shift forecasting, transfer learning 
\end{IEEEkeywords}

\section{Introduction}
Social media has become an integral part of hundreds of millions of people to perceive information and share it with others across the world in a short time. Although it provides an optimistic view on connecting people around the globe, social media forums affect core beliefs, values, and an attitude of its consumers with the type and quantity of the information it provides. Political bias, opinions, hate, and misinformation is more profound on current social media forums in recent years due to massive user participation and limited fact-checking on information that is shared among the general population. The prevalence of confirmation bias majorly affects user opinion on social media, which in turn makes online user communities to fall in the wide spectrum of political bias ranging from \emph{far-left} to \emph{far-right}~\cite{vicario2019polarization}. Political bias has been widely evidenced in recent events like COVID vaccine hesitancy~\cite{cossard2020falling} and gun-control. Studying political bias in social media can have direct implications over several research areas like hate speech detection, misinformation, and echo chambers modeling as all such problems would involve a large volume of biased conversations on social media.

Machine learning, with advancements in natural language processing and deep learning, has been actively used in studying political bias on social media. But the key challenge to model political bias is the requirement of human effort to label the seed social media posts to train machine learning models. Although very effective, this approach has disadvantages in the time-consuming data labeling process and the cost to label significant data for machine learning models is significantly higher. The web offers invaluable data on political bias starting from biased news media outlets publishing articles on socio-political issues to biased user discussions about several topics in multiple social forums. In this work, we introduce a novel approach to label political bias for social media posts directly from US congressional speeches without any human intervention for downstream machine learning models. Also, existing works model political bias as a prediction problem to predict the political leaning of given users, news articles, or social media posts. However, political bias in social media can shift as user opinion on topics change over time. Forecasting the political bias of given topics or conversations can have advantages in foreseeing user opinion on social media. In this work, we analyze a diverse set of features collected from social media text, users, sentiment, and cascades for traditional machine learning models to examine political bias in both prediction and forecasting aspects.



\begin{figure}[t!]
    \centering
    \subfloat[\centering Frequent Twitter Entities]{{\includegraphics[scale=0.4]{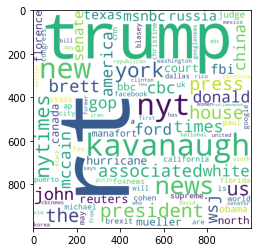}}}
    \qquad
    \subfloat[\centering Frequent Gab Entities]{{\includegraphics[scale=0.4]{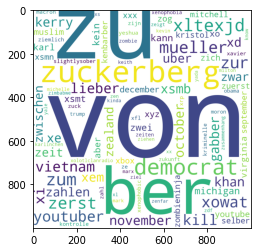}}}
    \caption{Frequent entities in Twitter and Gab posts from January 2018 to September 2018. Wide range of entities used in our work instead of targetting keywords for a specific problem.}
    \label{fig:Entities}
\end{figure}

In this work, we use posts from two popular social media forums: \emph{Twitter} and \emph{Gab}, which have their own set of political ideologies on free speech. Gab is relatively a newer social media forum that emphasizes on \emph{free speech} to often align with far-right groups' ideology. Gab is the platform of interest in computational social science research communities to study hate speech, far-right echo chambers, and racism. Twitter on the other hand makes restrictions on tweets, news articles, and user accounts that glorify hate to limit the dissemination of hate speech and misinformation in social media. We use publicly available Gab~\cite{fair2019shouting} and Twitter~\cite{brena2019news} datasets aiming to design the best feature space for both tasks.

In this work, we study multiple aspects of political ideology and bias in posts collected from two social media forums that are diverse in ideology. We exclusively utilize openly available text data such as real speeches and debates collected from politicians as surrogates to study the political bias. In particular, our contributions are three-fold in this work: \begin{enumerate}
	\item We present a method that gives political bias score labels of social media posts by adapting representations of entities learned from congressional speeches
	\item We give a machine learning approach to the predict political bias of posts using text embeddings. We extend our study and show that the model learned from one social media (Twitter) can be transferred to another ideologically distinct forum (Gab)    
    \item We provide a list of features, including engineered features like linguistic, cascade, and user features, along with contextual text embeddings, for machine learning models to forecast the political bias of a conversation
\end{enumerate}


\section{Related Work}

Social media produces large amounts of data that can be exploited to extract valuable information containing human interactions and opinions. There is a rising concern in recent years that such social media forums can be responsible for causing political bias in people which has the potential in affecting presidential elections~\cite{9026882} and news consumption~\cite{Garimella2021PoliticalPI}. Political bias detection from the given text using machine learning approaches has been a growing interest among researchers. Most of the early approaches detect bias using a traditional "bag-of-words" based classifier to focus on word lexicons~\cite{10.5555/3104482.3104544}. The major obstacle of these approaches is it heavily relies on primary-level lexical information and neglects semantic structure. Several studies have tried to detect bias on the multitude of data formats inducing words~\cite{Spinde2021AutomatedIO}, sentences~\cite{10.1145/3184558.3191640}, articles~\cite{chen-etal-2018-learning}, and news mediums~\cite{baly-etal-2018-predicting}.

Neural network based approaches have been widely used in political bias detection and ideology detection problems in past years. Recurrent Neural Network (RNN) has been used to accumulate the political leaning of each word to determine sentence level bias~\cite{iyyer2014political}.  RNN has also been used in determining political bias in multiple levels such as word, paragraph, and also at discourse level~\cite{chen-etal-2020-analyzing}. It is well established that news articles can have the political bias of left-leaning or right-leaning. The work in~\cite{chen-etal-2018-learning} provided a method to generate an article with the same topic but with flipped political leaning using NLP methods. Attention-based multi-view model focuses on sentences to identify corresponding political ideology~\cite{Kulkarni2018MultiviewMF}. Few other works study similar method at the sentence level in~\cite{Ji2017DistantSF} by extracting relations from text and news article headlines~\cite{gangula-etal-2019-detecting} along with several baseline models like SVM and CNN.

Other than the traditional bag-of-words based models, few works incorporated graphs to represent user opinions and ideologies in studying political bias. A recent work introduced an opinion-aware knowledge graph which infers based on circumstantial information from text and knowledge bases~\cite{10.5555/3172077.3172399} both by using previous background knowledge from the graph. TIMME framework~\cite{osti_10178656} handles heterogeneity of networks formed from social media sites. GCN~\cite{DBLP:conf/iclr/KipfW17} based approaches have also been introduced to encode social and textual information from news articles and social networks to capture political perspective~\cite{li-goldwasser-2019-encoding}. 


\begin{figure}[h]
    \centering
    \includegraphics[scale=0.35]{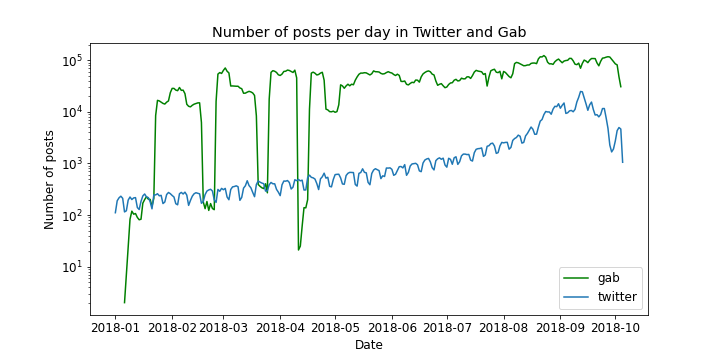}
    \caption{Number of posts per day in our Gab and Twitter data corpus. The rate of posts in Gab is high due to the availability of generic posts in Gab corpus unlike Twitter corpus, which is more focussed on politics tweets.}
    \label{fig:timeseries}
\end{figure}

\section{Datasets}
We utilize publicly available datasets from both social media and politicians' speeches in this study to study political bias.
 
\textbf{Congressional speeches}: Politics based data which are openly available on the web are good resources for machine learning models to learn bias makers concerning the given context like a topic, an event, or a person. Some useful resources include news articles, previously labeled politics-based social media posts, and actual political talks in US Congress and presidential debates. The political bias of news domains~\cite{adfontes} can divert the actual political leaning of the given topic mentioned in news articles. This may further introduce fake news which can prove disadvantageous for ML algorithms to map political bias. Similarly, a very limited set of labeled social media posts are easily accessible in recent years and are accurate in model training but difficult for ML models to generalize for the given problem in the given timeline. Political speeches, collected from politician's talks and debates, on the other hand, give a political party's actual viewpoints on topics, which can help in understanding the context with respect to all political parties. In this work, we utilize transcripts of congressional speeches~\cite{ontheissues} which consists of both republican and democratic speeches in congress. From a large corpus of congressional speeches, we used only \emph{1000} speeches from \emph{478} Democrat speeches and \emph{389} Republican speeches that align with topics discussed in social media posts.

\begin{figure}[h]
    \centering
    \includegraphics[scale=0.4]{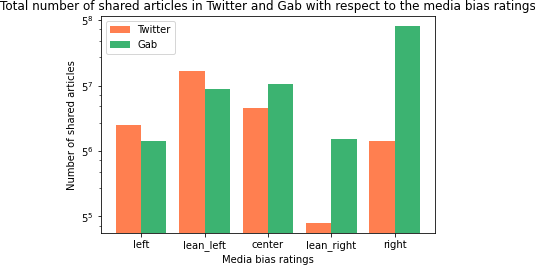}
    \caption{The number of news articles shared in Gab and Twitter data concerning their media bias shows Twitter leans \emph{left} and Gab supports \emph{right} politics. Note that our Twitter data corpus is significantly smaller than that of Gab.}
    \label{fig:barPlot}
\end{figure}

\textbf{Social media posts}: We use publicly available Twitter~\cite{brena2019news} and Gab~\cite{fair2019shouting} datasets in this work. The Twitter data~\cite{brena2019news} comprise of only posts that share news articles that discuss political topics from a selected news media sources. The dataset spans from January 2018 to September 2018 with a total of \emph{722,685} tweets. The dataset comprises news articles collected from news domains that range from a wide spectrum of political leaning and public tweets that mention such news articles. Our Gab dataset~\cite{fair2019shouting} comprises 40 million posts including all replies, re-posts, and quotes with URLs and hashtags that were submitted between 2016 and 2018. To make a fair analysis in this work, we only use the sampled Gab posts that are submitted between January 2018 and September 2018, same time range as the Twitter data. It is evident from Figure~\ref{fig:timeseries} that we have a large number of posts available in the Gab corpus. This is primarily because the Twitter data collected targeting the availability of politics based news articles, while the Gab data is not focused on any such domains. Gab is widely described as the social media which supports far-right ideologies. To verify this fact, we check the political bias of news media outlets, given by \emph{adfontesmedia}~\cite{adfontes}, that are shared in posts present in our data corpus and their corresponding frequency. Our summary is reported in Figure~\ref{fig:barPlot} which clearly shows Gab posts mainly share more news articles from \emph{far-right} and \emph{right-leaning} media outlets, while Twitter shares more news articles from \emph{left-leaning} media outlets.


\section{Methodology}
We present machine learning (ML)-based approaches for two studies performed on Twitter and Gab datasets described above. First, we propose an approach to label political bias score ($\gamma$) of social media posts and then we give approaches modeled as prediction and forecasting tasks to study political bias using the labeled data.

\subsection{Political bias score labeling}
\label{sect:labeling}
Machine learning models to predict political ideology or political bias of social media posts require a rich set of training data. Current methods rely on human annotations to obtain the training data which is time-consuming and specific to a defined problem. We present a method to label social media posts with political bias scores in the range $[-1,+1]$ where posts with the bias score of $-1$ represent \emph{far-left} and posts with bias score of $+1$ represent \emph{far-right}. Our proposed method to label social media posts consists of two characteristics: i) it is dependent on entities in general rather than on a specified problem, and ii) it relies on political party's (both republican and democrat) perspectives to get the context of entities. Our proposed method is more generic compared to existing studies in obtaining political bias as we use entities like \emph{hashtags}, \emph{person}, \emph{event}, and \emph{place} to label the data. We extract entities in both social media posts and congressional speeches after basic text pre-processing steps like case-folding, stop-words removal, and removing punctuations, with Stanford Named Entity Recognition (NER). In this work, we consider only common nouns and proper nouns as entities. We choose our congressional speeches data carefully to match entities available in either \emph{Gab} or \emph{Twitter} posts, so the entities extracted from social media posts would have context in one or both of the political parties. We consider entities from social media only if their occurrence frequency is at least 100 in at least one of the social media forums. Such entities are depicted in Figure~\ref{fig:Entities}, which illustrates the presence of multiple topics in both social media forums. 

We propose a method to label social media posts based on how the extracted entities are perceived by democrats and republicans. In this work, we utilize Term-Frequency Inverse Document Frequency (\emph{TF-IDF}) to get the importance of entities in terms of both republican and democrat perspectives. We consider that the political bias of a topic or entity varies according to the ideology of republicans and democrats. For a given social media post $S$, from Twitter or Gab, with a set of $n$ entities $E(S)=\{e_1,e_2, \ldots e_n\}$ where $n=[1,\infty)$, we propose to identify the political bias $\gamma(S)$ using Equation~\ref{eq:bias_calc}:

\begin{equation}
	\gamma(S) = \mathlarger{\mathlarger{\sum}}_{i=1}^n \frac{TF_r(e_i)-TF_d(e_i)}{TF_r(e_i)+TF_d(e_i)}
	\label{eq:bias_calc}
\end{equation}

where $TF_d(e_i)$ and $TF_r(e_i)$ are TF-IDF of the entity $e_i$ in the democrat and the republican context respectively. The above equation gives the political bias $\gamma(S) <= 0$, if entities $E(S)$ are supported more by democrats and $\gamma(S) > 0$, if republicans support $E(S)$.

\subsection{Feature engineering for political bias prediction}
\label{sect:prediction}
The availability of the data obtained using our method given in the previous section (Section~\ref{sect:labeling}) can provide multiple opportunities to train ML algorithms in several aspects. Since our labeling is not relevant to specific features in our datasets, we require to extract features from our text data for ML algorithms. In this work, we utilize only a very common set of stylistic features that are present in all types of text data along with the contextual text representations obtained using the \emph{FastText} model~\cite{joulin-etal-2017-bag}. FastText is an extension of the popular word representation framework \emph{Word2vec}, where the FastText extracts the contextual text representation by aggregating representations of the character n-grams. FastText proves to construct better word representations to even words that do not appear in the training corpus and boosts performance compared to its predecessor Word2vec. Our engineered set of features and their description are given in Table~\ref{table:prediction_features}. 

\begin{table}[h!]
\centering
\caption{Engineered features for political bias prediction task}
\begin{tabular}{||c | c||} 
 \hline
 \textbf{Feature} & \textbf{Description}  \\ [0.5ex] 
 \hline\hline
party & If democratic then 0 else 1 \\
n\_sen & Total sentences in a post \\
n\_word & Total words in a post \\
n\_char & Total character in a post \\
average sent. & Avg sentence length in the post \\
average word & Avg word length in the post \\
sentiment & Polarity score of a post \\
subjectivity & Subjectivity score of a post \\ [1ex]
 \hline
\end{tabular}
\label{table:prediction_features}
\end{table}

With the given set of features for a social media post $S$, we propose the political bias prediction as a binary classification problem $f: \gamma_x(S) \rightarrow 0|1$ where $\gamma_x(S)$ is the modified political bias score of social media post $S$. For simplicity purposes, we consider that $\gamma_x(S)=0$, if $\gamma(S) \leq 0$ and $\gamma_x(S)=1$, if $\gamma(S) > 0$ in this work.

\subsection{Forecasting political bias shift with information cascades}
\label{sect:forecast}
The political bias of user communities in online social media tend to change as users participate in conversations and respond in multiple activities like replying/commenting, liking, and re-sharing a post. Such shifts in the political bias of a topic or a conversation depends on the bias of users participating in conversations, the current sentiment of the conversation, and the context of topics in discussion.  Forecasting such shift in political bias can have a huge impact on understanding user opinion in social media, relating news media bias with user opinion, and identifying misinformation or fake news. In this work we propose political bias shift forecasting by utilizing information cascades and machine learning. Information cascades or \emph{Cascades} act as dynamical processes in social networks, which captures the complete evolution of a topic over time along with their virality and user interactions~\cite{zang2017quantifying}. In this work, we consider each conversation thread in the given social media as a cascade. The origin of a cascade begins with single-user post and then the user's social network slowly start a discussion about the post (in terms of replies and re-posts) along with their opinions. Thus, a cascade $\mathcal{C}$ can be represented as a graph: $\mathcal{C} = \{\mathcal{V}_t,\mathcal{M},z,w\}$, where $\mathcal{V}_t$ is a post/reply/re-post that appear at time step $t \in \{0,1, \ldots T\}$, $\mathcal{M}$ is a set of edges $\mathcal{M}=\{m_1,m_2, \ldots \}$ with $m_i: \mathcal{V}_{t_q} \rightarrow \mathcal{V}_{t_p}$ where $t_q > t_p$, $z$ is a node property representing sentiment score of post $v \in \mathcal{V}_t$, and $w$ is the edge weight of $m_i \in \mathcal{M}$ representing stance of post $\mathcal{V}_{t_q}$ about the post $\mathcal{V}_{t_p}$. Given a partial cascade $C_{j_r}$ and its corresponding bias score $\theta$ with only upto $d << T$ time steps, we propose the political bias shift as a binary classification problem: $g:\theta(C_{j_s}) \rightarrow 0|1$, where $s=d+1$. We propose to predict the presence/absence of bias shift at the $s^{th}$ time step in the cascade. We perform this study only on the Gab dataset as we do not possess complete conversation in the Twitter data. We engineer features based on multiple categories from the extracted cascades as given in Table~\ref{table:feat_cascades} for our proposed forecasting task. Along with engineered features we also use text representations of posts present in $C_{j_r}$.

\begin{table}[h!]
\caption{Multiple categories of features engineered from Gab corpus for bias forecasting.}
\centering
\begin{tabular} { | p {2 cm}| p {1.2 cm}| p {4.2 cm}|}
\hline
Category & Total features & Example \& Description \\
\hline
\textbf{User} & 8 & Root Influence: Neighbors of root divided by number of followers of root node, Border Influence: Number of followers that are not in cascade\\
\hline
\textbf{Polarity} & 6 & Positive Influence: Average positive sentiment score of training nodes, Negative Influence: Average positive sentiment score of training nodes\\
\hline
\textbf{Cascade} & 11 & Sentiment reshares: Avg. no. of reshares of +ve sentiment posts - Avg. no. of reshares of -ve sentiment posts, Stance reshares: Avg. no. of reshares of +ve stance posts - Avg. no. of reshares of -ve stance posts\\
\hline
\textbf{Temporal} & 5 & Root activity: Avg. time gap between posts by the root user upto the current cascade, Root response: Avg. response time of root user upto the current post\\
\hline
\end{tabular}
\label{table:feat_cascades}
\end{table}

\section{Results}
We examine and validate our proposed methods in three possible ways: 

\begin{enumerate}
	\item \textbf{Validation}: We validate our data labeling by comparing the overall labeled political bias of social media forums with similarity measures in contextual text representations
	\item \textbf{Prediction Analysis}: We give multiple quantitative methods for political bias prediction using traditional ML models with features described in Section~\ref{sect:prediction}
	\item \textbf{Forecasting Analysis}: With cascades from the \emph{Gab} dataset, we give a quantitative evaluation on ML models to forecast political bias shift in conversations with features described in Section~\ref{sect:forecast}
\end{enumerate}

\subsection{Data label validation}
The entity extraction step in our data labeling method identified \emph{131,345} and \emph{78,546} unique political entities in Twitter and GAB data respectively. We observed that more than \emph{20,154} entities are common in both platforms. The word cloud for frequent entities for both platforms are given in Figure~\ref{fig:Entities}. We validate our data labeled with a political bias score labeled using the method described in Section~\ref{sect:labeling} using quantitative reasoning in both Twitter and Gab. In particular, we compare contextual text representations of social media posts with contextualized representations of democrat and republican speech transcripts. 

\begin{figure}[h]
    \centering
    \includegraphics[scale=0.4]{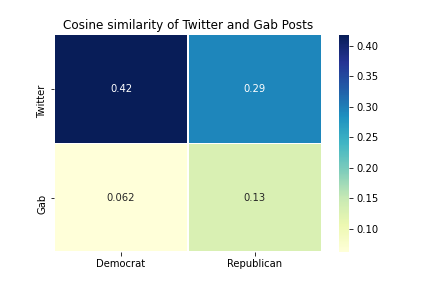}
    \caption{Cosine similarity scores based on text representations from Twitter and Gab aligns with Democrat and Republican transcripts respectively}
    \label{fig:cosine similarity}
\end{figure}

To validate our data labels, we used cosine similarity to compare contextual text representations of social media posts with transcripts. To obtain the contextual representations we utilize a pre-trained \emph{FastText} model for both congressional speech transcripts and social media posts. To compute overall political bias vectors of democrat/republican transcripts and Twitter/Gab posts by using an element-wise mean of the corresponding vectors. For example, we get political bias vector of Twitter by using an element-wise mean of vectors from all Twitter posts. With political bias vectors of all datasets, we compute cosine similarity of each pair and we give results in Figure~\ref{fig:cosine similarity}. Our results align with a simple data summary given in Figure~\ref{fig:barPlot} based on comparing news articles shared in each social media forum. We easily can evidence that overall Twitter posts are \emph{left-leaning} and Gab posts are \emph{right-leaning}. The higher similarity measures also approximately correlate with the mean bias scores of posts given in Table~\ref{table:mean_label}, which is calculated completely from our proposed labeling method. The similarity scores of Gab are much lower than the Twitter. This is due to noisy Gab posts in our corpus. Twitter data is more focused on politics based posts, whereas Gab posts are combination of general and political posts.


\begin{table}[h!]
\centering
\caption{Summary of political bias scores in Twitter and GAB posts labeled by the proposed method}
\begin{tabular}{||c | c||} 
 \hline
 \textbf{Measure} & \textbf{Score}  \\ [1ex] 
 \hline\hline
Sum of bias scores for GAB & 15.65468246877 \\
Mean of bias scores for GAB & 0.101471913141 \\
Median of bias scores for GAB & 0.088956416627 \\
Sum of bias scores for Twitter & -50.48289202612 \\
Mean of bias scores for Twitter & -0.394802532272 \\
Median of bias scores for Twitter & -0.278794265613 \\
 \hline
\end{tabular}
\label{table:mean_label}
\end{table}



\vspace{-0.5cm}
\subsection{Political bias prediction}
With the labeled social media data available for ML models, we perform two analyses with engineered and contextual features mentioned in Section~\ref{sect:prediction}. Both analyses are highly motivated by transfer learning models, which train on a dataset collected from one domain without seeing the test data which is from another domain. Also, we use four traditional ML models: \emph{Random Forests}, \emph{Multi Layer Perceptron}, \emph{Decision Trees}, and \emph{Linear Regression} for both analyses. We evaluate all models in terms of Accuracy, Precision, Recall, F-Score, and AUROC measures. 

In the first analysis, we use the congressional transcripts as the training data and posts from each social media as the test set. The results are summarized in Table~\ref{table:transfer1}. The features that we extract help ML algorithms to predict political bias of both social media posts without even a fraction of social media posts available for training. In particular, models performance in Twitter data is comparatively better than that of Gab as the Twitter data is more focused on politics unlike the Gab data. Overall, the performance of Linear Regression is higher (with a minimum of $7\%$ increase) than MLP and Decision Trees in terms of Recall, F-Score, and AUROC measures in Twitter, while the Random Forests give competing performance with Linear Regression. Model performances in Gab is close between Linear Regression, MLP, and Random Forests with Linear Regression performing $1\%$ higher/lower compared to the MLP.


\begin{table}[h!]
\centering
\caption{ML models performance ($\%$) with transcripts as training set and social media posts as test set}
\begin{tabular} { |p {1.2 cm}|p {1.2 cm}|p {1.2 cm}|p {1.2 cm}|p {1.2 cm}|}
\hline
\multicolumn{5} { | c | }{Twitter Data}\\
\hline
---- & RF & MLP & DT & LR \\
\hline
Accuracy & \textbf{0.704878} & 0.643902 & 0.626829 & \textbf{0.704878}\\
Precision & \textbf{0.725274} & 0.674846 & 0.631578 & 0.713541\\
Recall & 0.650246 & 0.541871 & 0.591133 & \textbf{0.674876}\\
FScore & 0.685714 & 0.601092 & 0.610687 & \textbf{0.693670}\\
ROC & \textbf{0.720937} & 0.660360 & 0.626484 & 0.699079\\
\hline
\hline
\multicolumn{5} { | c | }{GAB Data}\\
\hline
---- & RF & MLP & DT & LR \\
\hline
Accuracy & 0.643902 & \textbf{0.651219} & 0.541463 & 0.648780\\
Precision & 0.624309 & \textbf{0.627659} & 0.507614 & 0.623036\\
Recall & 0.591623 & 0.617801 & 0.523560 & \textbf{0.623036}\\
FScore & 0.607526 & 0.622691 & 0.515463 & \textbf{0.623036}\\
ROC & 0.664275 & 0.650553 & 0.540318 & \textbf{0.668387}\\
\hline
\end{tabular}
\label{table:transfer1}
\end{table}

In the second analysis, we completely ignore the transcripts and compare the performance of ML models trained on one social media without the knowledge of another dataset. Results have been summarized in Tables~\ref{table:gab_twitter} and ~\ref{table:twitter_gab}. Based on given results, we can see that the models trained on Gab posts transfer well on the Twitter data with Linear regression achieving at least $60\%$ in all the measures. Again, the models trained on Twitter could not give the comparative performance while tested on Gab posts with all ML models achieving around $55\%$ in all measures. This again validates that models trained on generic data like Gab can effectively transfer to more focused datasets like Twitter, but not the other way.

\begin{table}[h!]
\centering
\caption{ML models performance ($\%$) with GAB posts as training set and Twitter posts as test set}
\begin{tabular} { |p {1.2 cm}|p {1.2 cm}|p {1.2 cm}|p {1.2 cm}|p {1.2 cm}|}
\hline
\multicolumn{5} { | c | }{Twitter Data}\\
\hline
---- & RF & MLP & DT & LR \\
\hline
Accuracy & 0.555667 & 0.540667 & 0.524667 & \textbf{0.611667}\\
Precision & 0.563688 & 0.564516 & 0.528387 & \textbf{0.605566}\\
Recall & 0.529062 & 0.392999 & 0.540951 & \textbf{0.661162}\\
FScore & 0.545826 & 0.463396 & 0.534595 & \textbf{0.632144}\\
ROC & \textbf{0.578461} & 0.578461 & 0.578461 & 0.578461\\
\hline
\end{tabular}
\label{table:gab_twitter}
\end{table}

\begin{table}[h!]
\centering
\caption{ML models performance ($\%$) with Twitter posts as training set and GAB posts as test set}
\begin{tabular} { |p {1.2 cm}|p {1.2 cm}|p {1.2 cm}|p {1.2 cm}|p {1.2 cm}|}
\hline
\multicolumn{5} { | c | }{GAB Data}\\
\hline
---- & RF & MLP & DT & LR \\
\hline
Accuracy & \textbf{0.555333} & 0.531333 & 0.536667 & \textbf{0.553333}\\
Precision & \textbf{0.563661} & 0.534472 & 0.543318 & 0.556254\\
Recall & 0.539171 & 0.576695 & 0.532587 & \textbf{0.58262}\\
FScore & 0.551144 & 0.554782 & 0.537899 & \textbf{0.569132}\\
ROC & \textbf{0.559660} & 0.55966 & 0.55966 & 0.55966\\
\hline
\end{tabular}
\label{table:twitter_gab}
\end{table}

\vspace{-0.5cm}
\subsection{Political bias shift forecasting}
\begin{figure}[h]
    \centering
    \includegraphics[scale=0.3]{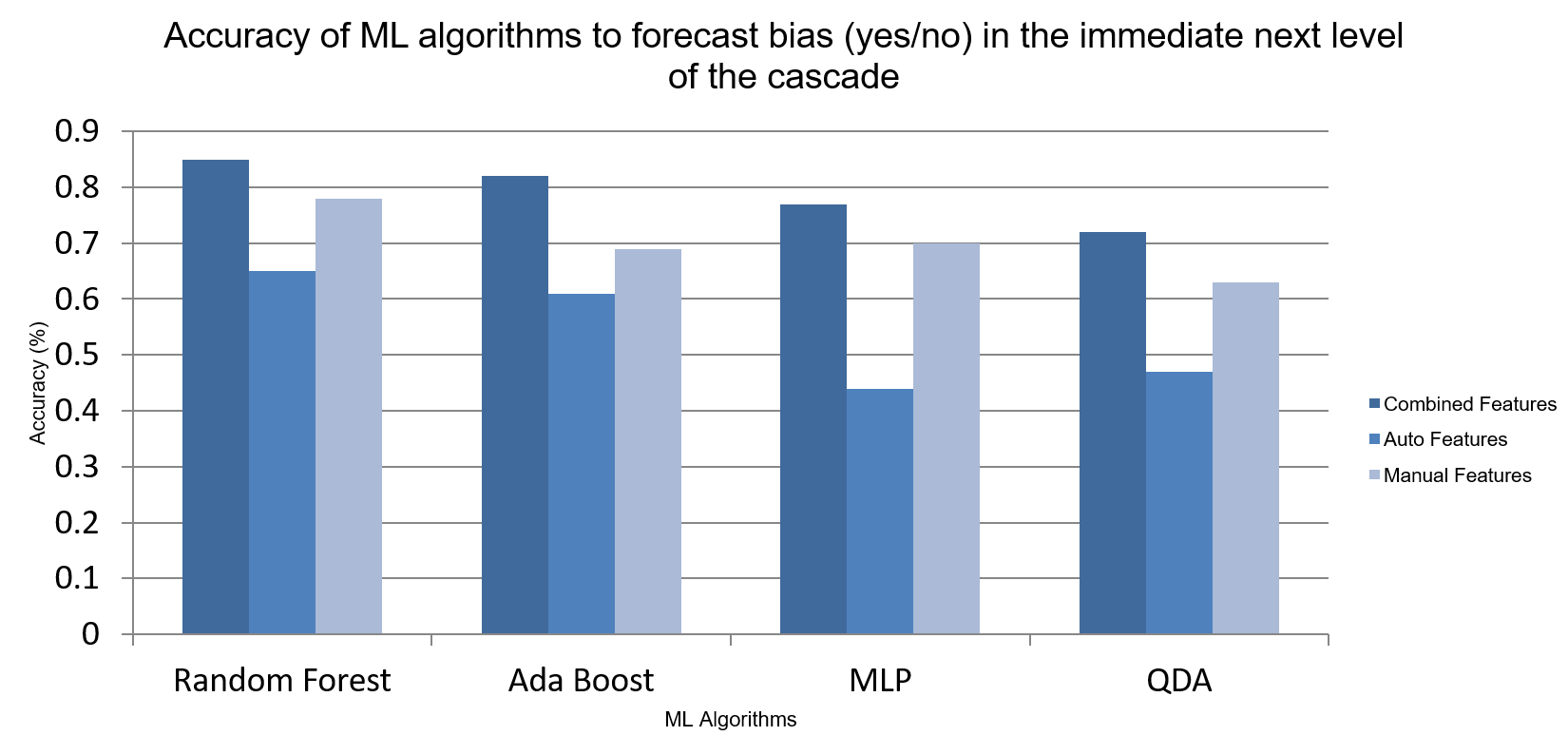}
    \caption{Accuracy ($\%$) of ML models to forecast polarity bias shift in Gab cascades with multiple feature sets}
    \label{fig:polarity_change_prediction}
\end{figure}

As mentioned earlier, we conduct this study only on the Gab dataset due to the availability of cascades in the Gab data. In total, we collected \emph{3.6 million} cascades and out of which we considered only \emph{69,746} cascades that have a minimum 5 levels. Our constraint on cascades with 5 levels is completely arbitrary in this work. As mentioned in Section~\ref{sect:forecast}, we used both engineered features and text features of cascade posts from the FastText model. In each cascade, we used nodes until an arbitrary number of levels $l > 5$ for training and we used ML models to predict if there is a bias shift in the $(l+1)^{th}$ level of the cascade. We present our forecasting results with engineered, auto-extracted, and combined features in Figure~\ref{fig:polarity_change_prediction}. We used four ML models: Random Forests, Ada Boost, MLP, and Quadratic Discriminant Analysis (QDA) in this study. 

\vspace{-0.5cm}
\section{Conclusion and Future work}
In this paper, we provided two approaches to study political bias in online social media forums. We presented a methodology to get political bias of social media posts using congressional speeches which contain politicians' perspectives on entities present in the posts. With publicly available large-scale social media posts from Gab and Twitter we provided a list of features for ML models to predict the political bias scores. We also engineered multiple categories of features from information cascades in Gab conversations for the bias forecasting task. Our multi-dimensional quantitative evaluation showed that our extracted set of features can help prediction and forecasting tasks profoundly.

Our proposed data labeling procedure is very generic such that it can be applied to multiple perspectives to study problems like media bias and fake news in the future. Even though the given method is more effective, more thorough case studies with valid ground-truth data can help to understand and update the approach for more robust political bias modeling in social media. The entities considered in this work can also be extended to include a cleaner and richer set of entities to overcome noise in the data. Another future scope is to explore on graph and text representation learning methodologies for features to use in the prediction or forecasting models.




\bibliographystyle{IEEEtran}  
\bibliography{bibliography} 

\begin{thebibliography}{10}
\providecommand{\url}[1]{#1}
\csname url@samestyle\endcsname
\providecommand{\newblock}{\relax}
\providecommand{\bibinfo}[2]{#2}
\providecommand{\BIBentrySTDinterwordspacing}{\spaceskip=0pt\relax}
\providecommand{\BIBentryALTinterwordstretchfactor}{4}
\providecommand{\BIBentryALTinterwordspacing}{\spaceskip=\fontdimen2\font plus
\BIBentryALTinterwordstretchfactor\fontdimen3\font minus
  \fontdimen4\font\relax}
\providecommand{\BIBforeignlanguage}[2]{{%
\expandafter\ifx\csname l@#1\endcsname\relax
\typeout{** WARNING: IEEEtran.bst: No hyphenation pattern has been}%
\typeout{** loaded for the language `#1'. Using the pattern for}%
\typeout{** the default language instead.}%
\else
\language=\csname l@#1\endcsname
\fi
#2}}
\providecommand{\BIBdecl}{\relax}
\BIBdecl

\bibitem{vicario2019polarization}
M.~D. Vicario, W.~Quattrociocchi, A.~Scala, and F.~Zollo, ``Polarization and
  fake news: Early warning of potential misinformation targets,'' \emph{ACM
  Transactions on the Web (TWEB)}, vol.~13, no.~2, pp. 1--22, 2019.

\bibitem{cossard2020falling}
A.~Cossard, G.~D.~F. Morales, K.~Kalimeri, Y.~Mejova, D.~Paolotti, and
  M.~Starnini, ``Falling into the echo chamber: the italian vaccination debate
  on twitter,'' in \emph{AAAI ICWSM}, 2020, pp. 130--140.

\bibitem{fair2019shouting}
G.~Fair and R.~Wesslen, ``Shouting into the void: A database of the alternative
  social media platform gab,'' in \emph{AAAI ICWSM}, vol.~13, 2019, pp.
  608--610.

\bibitem{brena2019news}
G.~Brena, M.~Brambilla, S.~Ceri, M.~Di~Giovanni, F.~Pierri, and G.~Ramponi,
  ``News sharing user behaviour on twitter: a comprehensive data collection of
  news articles and social interactions,'' in \emph{AAAI ICWSM}, vol.~13, 2019,
  pp. 592--597.

\bibitem{9026882}
L.~Belcastro, R.~Cantini, F.~Marozzo, D.~Talia, and P.~Trunfio, ``Learning
  political polarization on social media using neural networks,'' \emph{IEEE
  Access}, vol.~8, pp. 47\,177--47\,187, 2020.

\bibitem{Garimella2021PoliticalPI}
V.~Garimella, T.~Smith, R.~Weiss, and R.~West, ``Political polarization in
  online news consumption,'' in \emph{ICWSM}, 2021, pp. 152--162.

\bibitem{10.5555/3104482.3104544}
S.~M. Gerrish and D.~M. Blei, ``Predicting legislative roll calls from text,''
  in \emph{ICML}, 2011, p. 489–496.

\bibitem{Spinde2021AutomatedIO}
T.~Spinde, L.~Rudnitckaia, J.~Mitrovic, F.~Hamborg, ichael Granitzer, B.~Gipp,
  and K.~Donnay, ``Automated identification of bias inducing words in news
  articles using linguistic and context-oriented features,'' \emph{Inf.
  Process. Manag.}, vol.~58, p. 102505, 2021.

\bibitem{10.1145/3184558.3191640}
C.~Hube and B.~Fetahu, ``Detecting biased statements in wikipedia.''\hskip 1em
  plus 0.5em minus 0.4em\relax WWW, 2018, p. 1779–1786.

\bibitem{chen-etal-2018-learning}
W.-F. Chen, H.~Wachsmuth, K.~Al-Khatib, and B.~Stein, ``Learning to flip the
  bias of news headlines,'' in \emph{11th ACL International Conference on
  Natural Language Generation}, 2018, pp. 79--88.

\bibitem{baly-etal-2018-predicting}
R.~Baly, G.~Karadzhov, D.~Alexandrov, J.~Glass, and P.~Nakov, ``Predicting
  factuality of reporting and bias of news media sources.''\hskip 1em plus
  0.5em minus 0.4em\relax ACL, 2018, pp. 3528--3539.

\bibitem{iyyer2014political}
M.~Iyyer, P.~Enns, J.~Boyd-Graber, and P.~Resnik, ``Political ideology
  detection using recursive neural networks,'' in \emph{ACL}, 2014, pp.
  1113--1122.

\bibitem{chen-etal-2020-analyzing}
W.-F. Chen, K.~Al~Khatib, H.~Wachsmuth, and B.~Stein, ``Analyzing political
  bias and unfairness in news articles at different levels of granularity,'' in
  \emph{Fourth ACL Workshop on Natural Language Processing and Computational
  Social Science}, 2020, pp. 149--154.

\bibitem{Kulkarni2018MultiviewMF}
V.~Kulkarni, J.~Ye, S.~Skiena, and W.~Y. Wang, ``Multi-view models for
  political ideology detection of news articles,'' in \emph{EMNLP}, 2018, pp.
  3518--3527.

\bibitem{Ji2017DistantSF}
G.~Ji, K.~Liu, S.~He, and J.~Zhao, ``Distant supervision for relation
  extraction with sentence-level attention and entity descriptions,'' in
  \emph{AAAI}, vol.~31, no.~1, 2017.

\bibitem{gangula-etal-2019-detecting}
R.~R.~R. Gangula, S.~R. Duggenpudi, and R.~Mamidi, ``Detecting political bias
  in news articles using headline attention,'' in \emph{ACL Workshop
  BlackboxNLP: Analyzing and Interpreting Neural Networks for NLP}, 2019, pp.
  77--84.

\bibitem{10.5555/3172077.3172399}
W.~Chen, X.~Zhang, T.~Wang, B.~Yang, and Y.~Li, ``Opinion-aware knowledge graph
  for political ideology detection,'' in \emph{IJCAI'17}.\hskip 1em plus 0.5em
  minus 0.4em\relax AAAI Press, pp. 3647--3653.

\bibitem{osti_10178656}
Z.~Xiao, W.~Song, H.~Xu, Z.~Ren, and Y.~Sun, ``Timme: Twitter
  ideology-detection via multi-task multi-relational embedding,'' \emph{ACM
  SIGKDD}, pp. 2258--2268, 2020.

\bibitem{DBLP:conf/iclr/KipfW17}
T.~N. Kipf and M.~Welling, ``Semi-supervised classification with graph
  convolutional networks,'' in \emph{5th International Conference on Learning
  Representations, {ICLR} 2017}, 2017.

\bibitem{li-goldwasser-2019-encoding}
C.~Li and D.~Goldwasser, ``Encoding social information with graph convolutional
  networks for{P}olitical perspective detection in news media,'' in \emph{ACL},
  2019, pp. 2594--2604.

\bibitem{adfontes}
``Media bias chart v6.0,'' \url{https://www.adfontesmedia.com/}, 2021, [Online;
  acessed 11-July-2021].

\bibitem{ontheissues}
``Ontheissues,'' \url{https://www.ontheissues.org}, 2021, [Online; accessed
  11-July-2021].

\bibitem{joulin-etal-2017-bag}
A.~Joulin, E.~Grave, P.~Bojanowski, and T.~Mikolov, ``Bag of tricks for
  efficient text classification,'' in \emph{Proceedings of the 15th Conference
  of the {E}uropean Chapter of the Association for Computational Linguistics:
  Volume 2, Short Papers}, 2017, pp. 427--431.

\bibitem{zang2017quantifying}
C.~Zang, P.~Cui, C.~Song, C.~Faloutsos, and W.~Zhu, ``Quantifying structural
  patterns of information cascades,'' in \emph{WWW}, 2017, pp. 867--868.

\end{thebibliography}



\end{document}